\documentclass{elsart}

\usepackage{epsfig}

\begin{document}

\begin{frontmatter}



\title{GAMMA: \underline{GAM}ma   \underline{M}ultiplicity Filter \underline{A}rray for the selection of angular momentum in heavy ion collision reactions}


\author[label1]{Deepak Pandit},
\author[label1]{S. Mukhopadhyay},
\author[label2]{Srijit Bhattacharya},
\author[label1]{Surajit Pal},
\author[label3]{A. De},
\author[label1]{S. R. Banerjee\corauthref{cor}}
\corauth[cor]{Corresponding author.}
\ead{srb@veccal.ernet.in}


\address[label1]{Variable Energy Cyclotron Centre, 1/AF-Bidhannagar, Kolkata-700064, India}
\address[label2]{Department of Physics, Darjeeling Government. College, Darjeeling-734101, India}
\address[label3]{Department of Physics, Raniganj Girls' College, Raniganj - 713347, India}

\begin{abstract}
A 50 element BaF$_{2}$ gamma-multiplicity filter has been designed, fabricated and installed successfully at the
Variable Energy Cyclotron Centre (VECC), Kolkata for angular momentum selection 
in heavy ion collision reactions. A thorough GEANT3 Monte Carlo simulation was performed to optimize the individual 
detector shape and geometrical structure of the array. The detectors have been fabricated in-house from bare barium fluoride crystals (each measuring 5 cm in length and having cross-sectional area of 3.5 x 3.5 cm$^2$). The basic detector properties have been studied using the lab standard $\gamma$-ray sources. The response function of the filter has been studied using GEANT3 for mapping the fold distribution of the filter onto the angular momentum space. The multiplicity filter has been used successfully to measure the high angular momentum of the compound nuclei 
populated in the reaction $^{20}$Ne (E = 145 $\&$ 160 MeV) + $^{93}$Nb, $^{27}$Al as well as low angular momentum
events from the fission fragments of $^{252}$Cf spontaneous fission source. The filter has also been used successfully for correcting non-statistical events involved in the heavy ion fusion reaction.   

\end{abstract}

\end{frontmatter}

\section{Introduction}
Several studies of the $\gamma$-decay of the giant dipole resonance in 
hot and fast rotating nuclei have focused on the various kinds of structure (triaxial, prolate, oblate, spherical) 
that the nuclear system can assume as it evolves towards the configuration of minimum energy 
and angular momentum \cite{Hara01, Gaard01, Snov01}. Unlike energy, the angular momentum involved 
in such interactions cannot be directly observed and are inferred only indirectly. Measurement of 
angular momentum on an event-by-event basis can provide unique information that can increase substantially 
our understanding of the complex nuclear interactions. The high spin states of hot compound nuclei lose most
of the excitation energy via particle and $\gamma$ emissions above the yrast line. The remainder of the excitation
energy and angular momentum is generally removed by the low energy yrast $\gamma$ emission \cite{Hara01}. 
A precise measurement of this $\gamma$-multiplicity is very important since the number of $\gamma$-rays emitted are directly related to the angular momentum populated in the system. 

A $\gamma$-multiplicity filter detector is a very important equipment for studying nuclear structure. 
Most of the country around the world have one form or the other \cite{Mat97, Jas83, Sam85, Nan03}.
Recently, a 50 - element gamma-multiplicity filter made of BaF$_{2}$ have also been designed and developed at VECC. 
The motivation behind fabricating this multiplicity filter is as follow:- 
\begin{enumerate}
	\item To measure the number of low energy discrete $\gamma$ rays in heavy ion reaction with large geometric                    efficiency.
	\item The detector should be very fast so that it can be used as a precise start trigger which will 
	compensate the relatively poor time structure of the cyclotron.
	\item It should be modular in nature so that different geometrical shape can be formed (castle/radial) as per the             requirement.
	\item It can be used to select the fusion events rejecting the non-fusion one.
\end{enumerate}

In this paper, we report the detailed simulation studies, detector fabrication procedure $\&$ their exhaustive testing using lab standard $\gamma$ ray sources and the complete performance of the filter array from very low to high spin
distribution measurement.

\section{Simulation}
\begin{figure}
\begin{center}
\includegraphics[height=6 cm, width=5 cm]{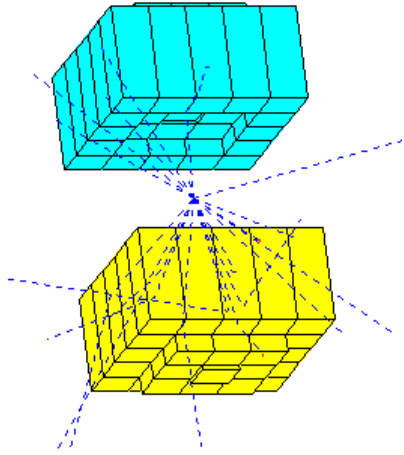}
\caption{\label{castle}Schematic view of the multiplicity filter array GAMMA in castle type geometry.}
\end{center}
\end{figure}

\begin{figure}
\begin{center}
\includegraphics[height=5 cm, width=7 cm]{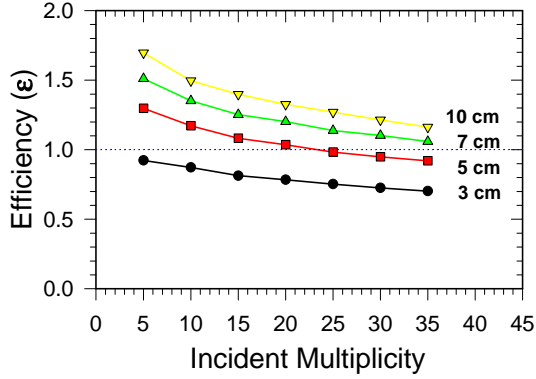}
\caption{\label{effi}Multiplicity detection efficiency vs. incident multiplicity plot for various 
detector lengths.}
\end{center}
\end{figure}

\begin{figure}
\begin{center}
\includegraphics[height=6 cm, width=8 cm]{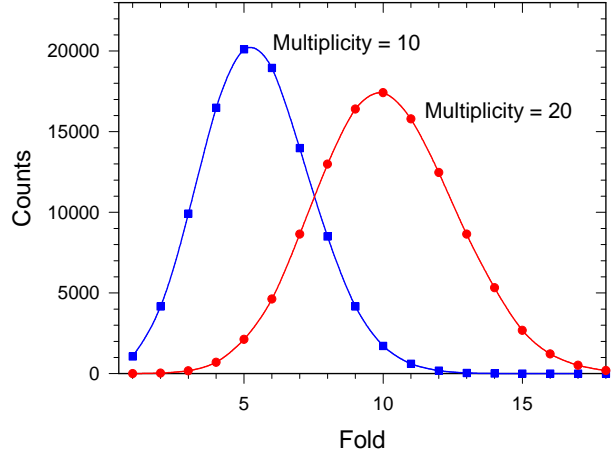}
\caption{\label{s_comp}Observed fold distribution for constant incident gamma multiplicities of 10 and 20.}
\end{center}
\end{figure}

A thorough GEANT3 Monte Carlo \cite{geant} simulation was done to select the optimum shape and geometrical structure of 
the multiplicity filter. The cross sectional area was kept same as the detector array LAMBDA  \cite{Supm} to make it granular and modular. To optimize the length of individual detector, simulation was 
performed considering the actual experimental scenario. In the realistic simulation, square shaped (cross-sectional area of 3.5 $\times$ 3.5 cm$^{2}$) detectors of different lengths (3, 5, 7 and 10 cm) made of BaF$_{2}$ material
were used. 50 detectors were arranged in two blocks of 25 elements each and placed 
at a distance of 7 cm on either side of the source. The central detectors were staggered in order to equalize the solid angle coverage. The schematic view of the setup as used in simulation is shown in Fig.\ref{castle}. Low energy gamma rays were thrown isotropically (Gaussian energy distribution varying from 0.3 to 1.5 MeV with peak at 0.7 MeV and 
FWHM is 0.5 MeV) with incident multiplicities of 5, 10, 15, 20, 25, 30, 35 and in each case, the peak position of
the fold distribution was analyzed.  It was found that the length of 5 cm 
has best possible average fold distribution (Fig.\ref{effi}) since its efficiency is close to 1 for a wide range of incident multiplicities. The fold detection efficiency ($\epsilon$ ) is defined as

\begin{equation}
\epsilon= \frac{M_{obs}}{M_{inc}*(\Omega/4\pi)}
\end{equation}
where, M$_{inc}$ being the incident multiplicity in 4$\pi$  geometry and M$_{obs}$ the observed multiplicity within 
solid angle   subtended by the gamma multiplicity filter. The fold distribution as observed for constant incident multiplicities of 10 and 20 for the detector having length of 5 cm is shown in Fig.\ref{s_comp}. 
The detectors of the multiplicity filter can be arranged in either castle or radial shape and simulation has been done for both the geometries. Although, castle type has larger cross talk probability than the radial type but the former has been preferred since in this geometry, detectors can be placed closer to the target and thereby subtending larger solid angle.

\section{Fabrication}
\begin{figure}
\begin{center}
\includegraphics[height=8 cm, width=7.0 cm]{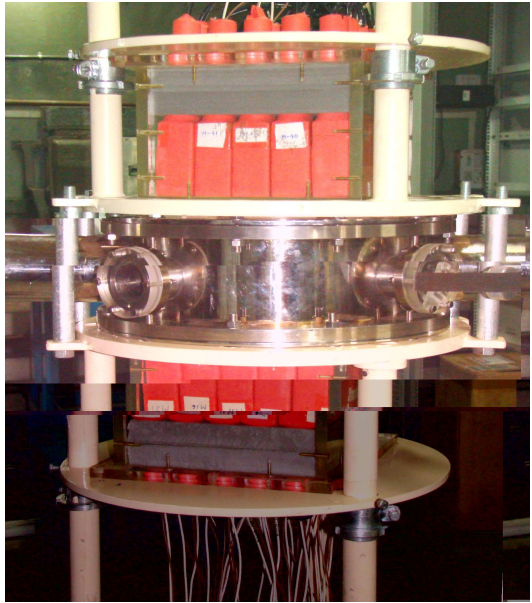}
\caption{\label{arr}Complete setup of the multiplicity filter array GAMMA kept on either side of the scattering chamber.}
\end{center}
\end{figure}

The BaF$_{2}$ crystals were procured from 
Beijing Glass and Research Institute, China. In general, BaF$_{2}$ has two scintillation light 
components; a fast component ($\tau$ = 0.6ns) peaking at $\lambda$ = 220 nm and a slow component ($\tau$ = 630ns) 
peaking at $\lambda$  = 320 nm with intensities 20$\%$ and 80$\%$, respectively. Standard procedures 
were followed for detector fabrication from bare barium fluoride crystals. First, the bare 
crystals were cleaned thoroughly using pure dehydrated ethyl alcohol. Each crystal  was 
wrapped with 6 layers of 15 $\mu$m white teflon tape since the scintillation light components 
are in the ultra violet (UV) region and teflon (C$_{2}$F$_{2}$) is high quality reflector of UV light. 
Next, aluminium foil of 10 $\mu$m (3-4 layers) was used to enhance the light 
collection and to block the surrounding light from entering into the crystal. Fast, UV sensitive 
photomultiplier tubes (29mm dia, Phillips XP2978) were coupled with the crystals using a highly 
viscous UV transmitting optical grease (Basylone, $\eta$ $\approx$ 300000 cstokes). This coupling need to be 
done very carefully so that no air bubble remains in the grease over the crystal surface, because 
air bubbles provide unwanted reflecting surfaces amounting to a loss of scintillation light which degrades the timing property. Specially designed aluminium collars were also used around the coupling area to provide additional support. A squared shape teflon reflector (35mm $\times$ 35mm) with a 30 mm hole at the centre was placed at the PMT end of the crystal to reflect back UV light which would otherwise escape 
from outside the PMT. A PMT voltage divider base was then attached to the PMT for applying the 
high voltage. After that, the whole assembly was covered with black electrical tape for light-tightness, 
and finally with heat-shrinkable PVC tube for providing mechanical stability to the detector.
A special scattering chamber and a detector stand have been designed, where the complete multiplicity
filter array can be kept at 3 cm  from the target position covering 71 $\%$  of
the total 4$\pi$ solid angle. The complete setup arranged in two blocks of 5 $\times$ 5 array, kept above and below   the scattering chamber at a distance of 5 cm from the target position is shown in Fig.\ref{arr}.

\section{Energy and Timing properties}
\begin{figure}
\begin{center}
\includegraphics[height=8 cm, width=7 cm]{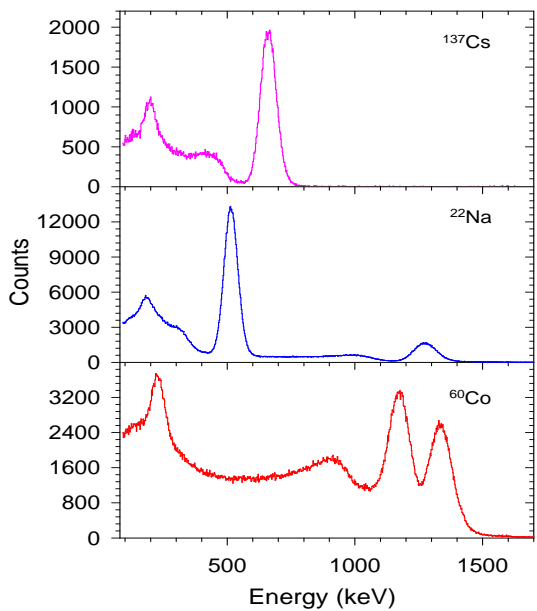}
\caption{\label{ener}Energy spectra from a single detector for different lab. standard gamma ray sources.}
\end{center}
\end{figure}

\begin{figure}
\begin{center}
\includegraphics[height=6 cm, width=7 cm]{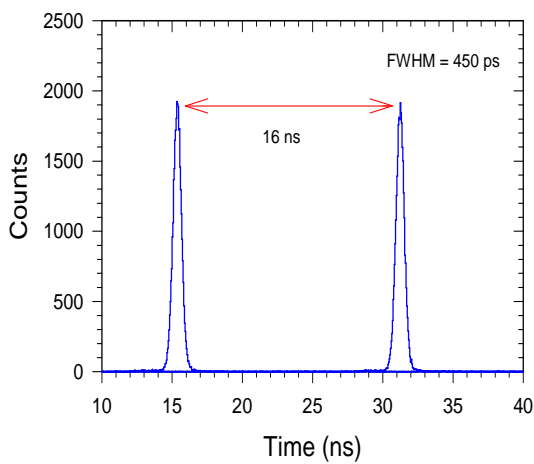}
\caption{\label{time}Time resolution of individual detector using $^{60}$Co $\gamma$-source.}
\end{center}
\end{figure}

The individual detector elements were tested with lab standard gamma ray sources. The energy 
spectra measured in individual detector is shown in Fig.\ref{ener}.  The time resolution between two BaF$_{2}$ 
detectors was  measured with the $^{60}$Co source. The source was placed in between two identical detectors 
which were kept 180$^{\circ}$ apart. The energies and their relative times were measured simultaneously in event 
by event mode. The energy gated  (1.0-1.4 MeV) time spectrum is shown in Fig.\ref{time}. The value obtained for 
time resolution is 450 ps. The cross talk probability was measured with $^{137}$Cs and $^{60}$Co sources. Nine 
detectors were arranged in a 3 $\times$ 3 matrix. A coincidence between the central detector and the rest of the 
eight sorrounding detectors was made to collect data for different hit patterns by counting in a scalar module. 
The ratio of the different hit to the single hit gave the cross-talk probability. The overall cross-talk probability of double hit was  found to be 2.5$\%$ $\&$ 15$\%$ for $^{137}$Cs $\&$ $^{60}$Co $\gamma$-sources respectively while that of triple hit was 0.04$\%$ and 0.37$\%$ respectively.

\section{Method for extraction of Angular Momentum}

In actual experiment, multiplicity filter registers folds which means number of low energy discrete
gamma rays detected in a single event. From this experimental fold, information about the populated angular momentum 
distribution of the compound nucleus can be extracted. A method has been developed in which the individual experimental fold is mapped onto the angular momentum space using the known 
response of the filter array generated by the simulation with GEANT. In this simulation, the 
experimental conditions are taken into account, including the detector thresholds and the 
trigger logic. The  $\gamma$ multiplicity (M$_\gamma$) of an event is related to the angular momentum (J) as  

\begin{equation}\label{rel}
J = 2M_\gamma + C
\end{equation}
where, C is the correction due to the angular momentum carried off by statistical gamma rays, evaporated particles and  the $\gamma$-rays below the trigger threshold. The J distribution can be obtained from statistical model code CASCADE \cite{cas}. The J and M$_\gamma$  distributions are triangular in nature with small diffuseness  and takes the form 

\begin{equation}\label{mul}
P(M_\gamma)=\frac{2M_\gamma+1}{1+exp[(M_\gamma-M_{max})/\delta m]}
\end{equation}
where, M$_{max}$ is the maximum of this distribution and $\delta$m is the diffuseness which are
adjusted for best fit between the experimental and the simulated fold distributions.

\begin{figure}
\begin{center}
\includegraphics[height=6 cm, width=8 cm]{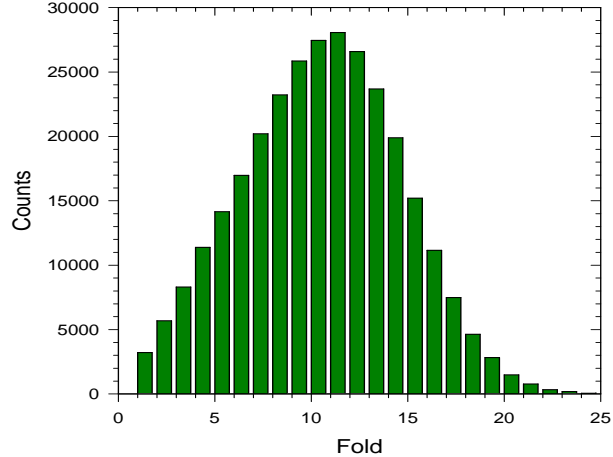}
\caption{\label{sim_f}Simulated fold distribution observed in the multiplicity filter.}
\end{center}
\end{figure}

\begin{figure}
\begin{center}
\includegraphics[height=6 cm, width=8 cm]{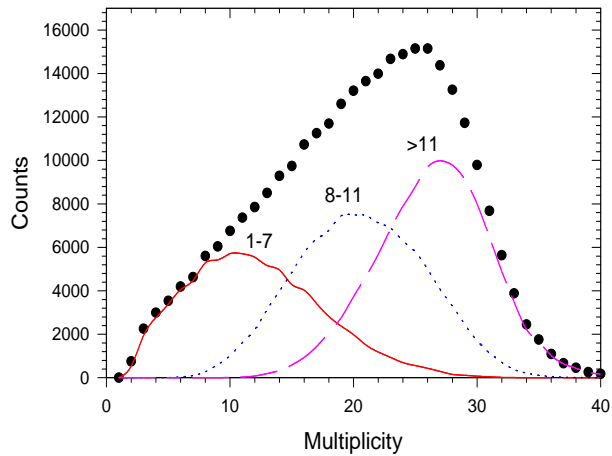}
\caption{\label{s_mult}The sampling of the total angular momentum distribution by gating on different
folds observed in the fold distribution of the multiplicity filter.}
\end{center}
\end{figure}

Low energy $\gamma$-rays having multiplicity distribution P$(M_\gamma)$ are made to incident on the 
filter array from the target centre isotropically and the fold distribution is recorded.
A fold distribution obtained for a setup of 50 detectors arranged in a castle type geometry is shown in Fig.\ref{sim_f}. Two blocks of 25 detectors arranged in 5 x 5 array were kept on either side of the source at a distance of 5 cm. The setup covered 56$\%$ of 4$\pi$ solid angle. 
The incident multiplicity distribution was taken as in 
eq. {\ref{mul}}, with M$_{max}$ = 30 and $\delta$m = 2, typical for 8 MeV/A reaction. The incident $\gamma$-energy distribution was taken gaussian with peak at 0.7 MeV and FWHM of 0.5 MeV. This simulated fold distribution is then compared with the experimentally observed fold distribution. The multiplicity distribution for different experimental 
folds can now be generated from the simulation and also the corresponding angular momentum distribution using 
eq. {\ref{rel}}.  The incident multiplicity distribution and the multiplicity distributions for different fold windows are shown in Fig.\ref{s_mult}, where the filled circle shows the incident multiplicity distribution, the continous line, dotted and the dashed lines indicate the multiplicity distributions gating on the events with folds 1-7, folds 8-11 and folds $>$ 11 respectively.

\section{Measurement of fold distributions from $^{252}$Cf and in-beam reaction }

A part of the GAMMA array has been used earlier for measuring high angular momentum in compound nuclear reactions. 
The detector array was tested in the experiment  $^{20}$Ne + $^{93}$Nb, at beam energies 145 and 160 MeV from K-130 AVF cyclotron at VECC \cite{Srij}. The multiplicity filter was configured in a castle geometry in
two blocks of 12 detectors each on either side of the target chamber at a distance of 10 cm from 
the target. The non-fusion events involved in the above fusion reactions were rejected using the observed fold
distribution of the multiplicity filter. The multiplicity filter has also been used for studying Jacobi shape transition of $^{47}$V populated in the reaction $^{20}$Ne + $^{27}$Al at beam energy of 160 MeV. The detailed setup and results of the experiment can be found in \cite{deep}.    

\begin{figure}
\begin{center}
\includegraphics[height=6 cm, width=8 cm]{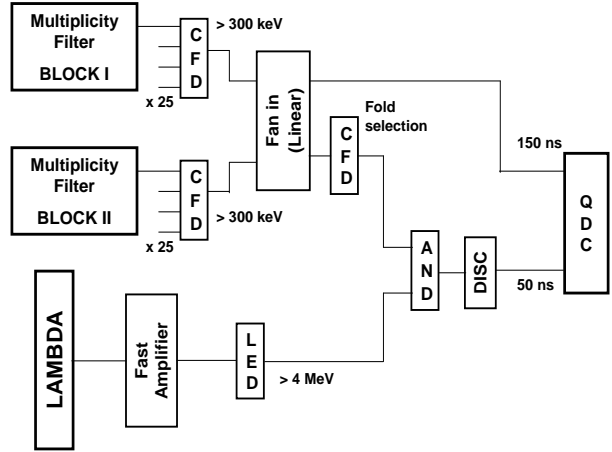}
\caption{\label{circuit}Circuit diagram of the experimental setup to obtain the fold distribution.}
\end{center}
\end{figure}

\begin{figure}
\begin{center}
\includegraphics[height=5 cm, width=7 cm]{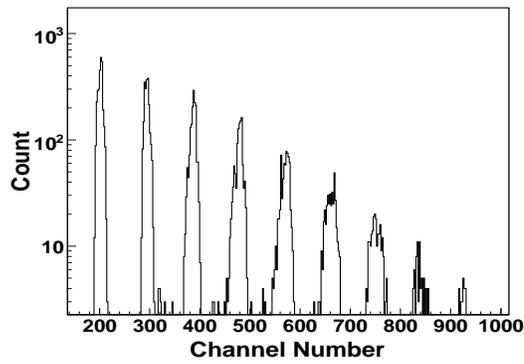}
\caption{\label{rfold}Experimental fold distribution showing folds $\geq$ 2 for $^{252}$Cf source.}
\end{center}
\end{figure}

Recently, the complete setup of 50 element multiplicity filter has been tested using a 3 $\mu$Ci $^{252}$Cf source for measuring low angular momentum events from the fission fragment. 
The multiplicity filter was arranged in two blocks of 5 $\times$ 5 array in castle type geometry. Each block was kept on either side of the source at a distance of 5 cm. The setup covers 56 $\%$ of 4$\pi$ solid angle. Along with 
the filter, a part of the LAMBDA (49 detectors arranged in 7 $\times$ 7) was also used to measure the high-energy gamma rays ($\geq$ 4 MeV) in coincidence with the low energy discrete $\gamma$-rays from the filter array. 
The output of the multiplcity filter is send to CAMAC CFD(C808, CAEN) to apply low threshold ($\geq$ 300 keV) for rejecting the low energy background.The SUM output of the CFD (50 mV for each hit) was then fed into a Linear FAN IN/OUT . The output of the FAN IN/OUT  was sent to another CFD for selecting the different fold conditions. Next, 
a coincidence between the output of the CFD and the high energy gamma ray ($\geq$ 4 MeV) coming 
from the main array was made to generate the gate signal for the QDC. Finally,
the output of the FAN IN/OUT was fed into the QDC for obtaining the fold distribution. 
The block diagram of the electronic scheme is shown in Fig.\ref{circuit}. The fold distribution for $^{252}$Cf source was generated applying the condition ($\geq$ 2 fold) and is shown in Fig.\ref{rfold}.  

\begin{figure}
\begin{center}
\includegraphics[height=5 cm, width=7 cm]{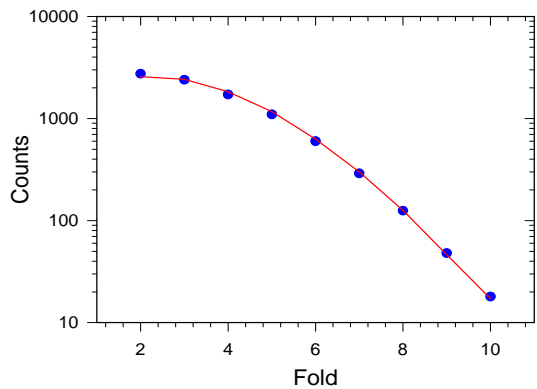}
\caption{\label{simfld}Experimental fold spectrum (symbols) fitted with GEANT simulation
(solid line) for $^{252}$Cf source. }
\end{center}
\end{figure}

\begin{figure}
\begin{center}
\includegraphics[height=6 cm, width=7 cm]{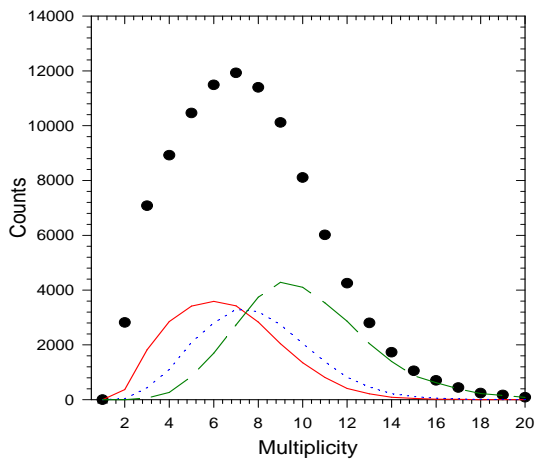}
\caption{\label{angu}The incident multiplicity distribution used in simulation (symbols). The multplicity distributions
obtained for different folds are also shown in the figure. The solid line represents fold 2, the dotted line represents
fold 3 and the dashed line represnets the multiplicity distribution for folds 4 and more.}
\end{center}
\end{figure}

The fold distribution obtained from the $^{252}$Cf source is mapped onto the angular momentum space using the 
response function generated by GEANT described earlier. The incident multiplicity 
distribution was considered triangular and the parameters M$_{max}$ and $\delta$m were found by fitting
the distribution obtained from \cite{Val01}. The M$_{max}$ and $\delta$m of the 
multiplicity distribution were found as 9 and 1.8. The energy distribution was considered as Gaussian with peak at 
0.65 MeV and width 0.75 MeV. The comparison between the experimental fold distribution and simulated 
fold distribution is shown in Fig.\ref{simfld}. The incident multiplicity distribution (symbols)  as well as the multiplicity distribution for different folds are shown in Fig.\ref{angu}. The average angular momenta 
for different fold distributions are given in table 1.

\begin{table}

\caption[]{\label{tab:nucl} Average angular momentum values corresponding
to different folds.}
\begin{center}		
\begin{tabular}{|c|c|c|}
\hline
Fold &	$\left\langle J \right\rangle$  & $\delta$J \\
 & $\hbar$ &  $\hbar$ \\
\hline
2 &     13.2    &    4.9    \\
\hline
3 &     15.8    &    5.1 	  \\
\hline
\,\,\,\, 4 or more \,\,\,\,  &  \,\,\,\,20.4 \,\,\,\,   & \,\,\,\,  6.6 \,\,\,\,  	\\
\hline
\end{tabular}
\end{center}		
\end{table}

\section{Summary}
A 50 element BaF$_{2}$ gamma-multiplicity filter has been designed, fabricated and installed successfully. The different detector properties have been studied using lab standard $\gamma$-ray sources. The filter has been 
used for the measurement of low energy discreate $\gamma$-rays from the reaction $^{20}$Ne + $^{93}$Nb,$^{27}$Al at beam energies 145 and 160 MeV as well as the fission fragments from the  $^{252}$Cf source. The GEANT3 Monte Carlo simulation has been performed cosidering realistic geometry to study the response function of the filter in order to map the fold distribution of the filter onto the angular momentum space. The filter has also been used for rejecting the non-fusion events involved in the heavy ion fusion reaction.



\end{document}